\documentclass[12pt]{article}
\usepackage{a4wide}
\usepackage{amssymb}
\usepackage{graphicx}
\usepackage{color}
\begin{document}
{\renewcommand{\thefootnote}{\fnsymbol{footnote}}
%\hfill  IGC--yy/m--n\\
%\medskip
\begin{center}
{\LARGE  Some implications of signature-change\\ in cosmological models of 
loop quantum gravity }\\ 
\vspace{1.5em}
Martin Bojowald$^{1}$\footnote{e-mail address: {\tt bojowald@gravity.psu.edu}}
and Jakub Mielczarek$^{2,3,4}$\footnote{e-mail address: {\tt jakub.mielczarek@uj.edu.pl}}
\\
\vspace{0.5em}
${^1}$Institute for Gravitation and the Cosmos, The Pennsylvania State University, \\
104 Davey Lab, University Park, PA 16802, USA\\
${^2}$Laboratoire de Physique Subatomique et de Cosmologie, UJF, INPG, CNRS, IN2P3
53, avenue des Martyrs, 38026 Grenoble cedex, France \\
${^3}$Institute of Physics, Jagiellonian University, {\L}ojasiewicza 11, 30-348 Cracow, Poland \\
${^4}$Department of Fundamental Research, National Centre for
Nuclear Research,\\ Ho{\.z}a 69, 00-681 Warsaw, Poland 
\vspace{0.5em}

\end{center}
}

\setcounter{footnote}{0}

\begin{abstract}
  Signature change at high density has been obtained as a possible consequence
  of deformed space-time structures in models of loop quantum gravity. This
  article provides a conceptual discussion of implications for cosmological
  scenarios, based on an application of mathematical results for mixed-type
  partial differential equations (the Tricomi problem). While the effective
  equations from which signature change has been derived are shown to be
  locally regular and therefore reliable, the underlying theory of loop
  quantum gravity may face several global problems in its semiclassical
  solutions.
\end{abstract}

\section{Introduction}

Anomaly-free effective models of loop quantum gravity, derived for spherically
symmetric configurations \cite{JR,HigherSpatial} and cosmological
perturbations at high density \cite{ScalarHol,ScalarHolInv}, have revealed an
unexpected phenomenon: At large curvature, signature change appears to be a
generic feature of quantum space-time geometry as provided by this theory
\cite{Action}. Not only the general phenomenon but also the specific form of
signature change seems to be universal in these models, giving further support
of the genericness of the effect. For the typical form of ``holonomy
modifications'' used widely in models of loop quantum gravity, the speed of
physical modes differs from the classical speed of light by a factor of
$\beta(h,K):=\alpha(h) \cos(2\delta K)$ where $\alpha(h)>0$ is a function of
the spatial geometry (the metric $h$), $\delta$ is a quantization parameter
often assumed to be related to the Planck length, and $K$ is a measure for
extrinsic curvature (or the Hubble parameter in cosmological models). For
large curvature $K>\pi/(4\delta)$, the speed is negative and commonly
hyperbolic mode equations turn elliptic, which in these models are of the form
\begin{equation} \label{Wave}
 \frac{\partial^2\phi}{\partial t^2}- \frac{\beta(h,K)}{a^2} \Delta\phi=S[\phi]
\end{equation}
with source and lower-derivative terms $S[\phi]$. All modes
$\phi$, gravitational as well as matter, are affected in the same way. 

The overall picture shows some relationships with other approaches to quantum
cosmology, mainly the no-boundary proposal of Hartle and Hawking
\cite{nobound}, and with other physical phenomena such as transonic flow or
phases of nano-wires. Related mathematical questions have been studied in the
mathematical literature since the 1930s, following seminal work by Tricomi
\cite{Tricomi}. Nevertheless, as a basis of effective space-time models,
equations of the form (\ref{Wave}) and the phenomenon of signature change show
several new and surprising features.  In this article, we give a general
presentation of implications in cosmology.

Section~\ref{s:Def} gives a definition of signature change in the absence of a
classical space-time metric, and contains a brief comparison with classical
models of signature change. (A review of the formal origin and gauge
independence of equation (\ref{Wave}) and signature change can be found in
App.~\ref{s:SpaceTime}.)  Section~\ref{s:Effects} provides several related
examples in different areas of physics together with the appropriate
mathematical formulation in terms of well-posed partial differential
equations. Section~\ref{s:Cosmo} applies these results to gravitational
questions, compares the pictures based on effective equations with
wave-function methods, draws cosmological implications, and ends with cautious
notes on global issues.

\section{Definitions}
\label{s:Def}

Despite first appearance, the wave equation (\ref{Wave}) is covariant under a
deformed algebra replacing classical coordinate or Poincar\'e
transformations. The corresponding quantum space-time structure is not
Riemannian, but has a well-defined canonical formulation using hypersurface
deformations, as reviewed briefly in the appendix.  

A Riemannian manifold has a covariant metric tensor which transforms by
$g_{a'b'}=(\partial x^a/\partial x^a{}') (\partial x^b/\partial x^b{}')
g_{ab}$ when coordinates are changed on the manifold. Accordingly, the line
element ${\rm d}s^2=g_{ab}{\rm d}x^a{\rm d}x^b=g_{a'b'}{\rm d}x^{a'}{\rm
  d}x^{b'}$ is invariant. Covariance under coordinate transformations (of
solutions to the field equations) is canonically represented as
gauge-invariance under hypersurface deformations in space-time
\cite{Regained}. Since the latter, along with Poincar\'e transformations, are
modified in the effective space-time structures we are considering, they can
no longer correspond to coordinate transformations. As a consequence, the
effective ``metric'' does not give rise to an invariant line element. Instead
of using metric-space notions such as geodesics and invariant scalar products
of 4-vectors, in order to extract predictions there is only the possibility of
computing canonical observables invariant under the modified gauge
transformations.  In this way, one still has access to all observable
information. However, the lack of a metric structure implies that many of the
convenient and well-known techniques of evaluating solutions of general
relativity are no longer available.

Although there is no standard notion of geodesics and light rays or null
lines, it remains straightforward to associate a causal structure to the
modified space-time structure underlying (\ref{Wave}), as long as
$\beta>0$. Instead of computing null lines for a metric, we just use
characteristics of the wave equation (\ref{Wave}), for instance for
gravitational waves to be specific. (In the presence of inverse-triad
corrections, scalar modes may propagate at speeds different from tensor modes
\cite{LoopMuk,ScalarHolInv}.) A characteristic is then a hypersurface which at
any point is normal to a vector $k^a$ satisfying $(k^t)^2-\beta
|\vec{k}|^2=0$. With these characteristics instead of null lines, we can
define light cones, a causal structure, and derived notions such as Penrose
diagrams.

Characteristics exist as long as $\beta>0$. For holonomy modifications,
$\beta$ changes sign at high density, and the characteristic equation no
longer has non-trivial solutions $k^a\not=0$. As a consequence, (\ref{Wave})
does not provide a causal structure in such a regime. Since the same condition
of $\beta<0$ makes the mode equation (\ref{Wave}) turn elliptic, we call this
regime ``Euclidean,'' while for $\beta>0$ we call it ``Lorentzian.'' In this
way, we generalize the two standard discrete choices of signature to a
continuous range of values taken by $\beta$ as it varies from the classical
limit $\beta=1$ at low curvature to a negative value at high curvature. Unless
$\beta=\pm1$, the canonical fields from which (\ref{Wave}) is derived do not
provide a classical notion of Lorentzian space-time or 4-dimensional Euclidean
space. But the most important properties regarding physical consequences,
including the existence of a causal structure and the type of initial or
boundary value problem required for reliable solutions, only depend on the
sign of $\beta$ rather than its precise value. For this reason, we still speak
of Euclidean or Lorentzian signature even if we have neither Euclidean nor
Lorentzian space(-time).

Signature change has been studied in quite some detail in classical general
relativity ($\beta=1$); see for instance \cite{SigChangeClass}. However, such
models are crucially different from what is considered here, in that they have
$\beta{\rm sgn}(\det h_{ab})$ (or $\beta{\rm sgn} N^2$) changing
discontinuously. As a consequence, in classical models the transition from
Euclidean to Lorentzian signature is always singular, which is not the case in
our effective models. (Note that for this reason signature change in the model
of \cite{SigChangeHybrid} is {\em not} analogous to deformed space-time
structures.) The subtleties and controversies related to the distributional
nature of solutions with classical signature change, discussed for instance in
\cite{ClassSigChange,ClassSigChange2,SigAbs,SigSmooth,SigChangeJunction,BoundarySigChange,GeometrySig},
do not play a role in our context. (The models we study could be considered as
a version of dynamical signature change as anticipated in the conclusions of
\cite{EuclLor}.)

It is necessary to consider inhomogeneity in order to see modified space-time
structures in which modes obey (\ref{Wave}). In particular the phenomenon of
signature change, which is the main topic of this article and is realized
because $\beta$ may change sign, can only be seen in inhomogeneous models: it
is manifested by the relative sign in space and time derivatives in field
equations. However, signature change (or a modified space-time structure) is
not a consequence of inhomogeneity, which in the perturbative context would be
dubious \cite{DeformedCosmo}. The origin of signature change lies in the
modification that one makes in the classical dynamics if one quantizes the
theory following loop quantum gravity, which provides operators for holonomies
(exponentiated and integrated connections) instead of ordinary connection
components. This modification appears already in the background dynamics of a
homogeneous minisuperspace model, but in this context one does not notice
signature change because all spatial derivatives vanish. Perturbative
inhomogeneity then makes the effect visible. Still, inhomogeneity or
perturbation theory is not the origin of signature change, which is easy to
see by the following two arguments: First, signature change persists in a
perturbative field equation no matter how small the inhomogeneity is, as long
as it is non-zero. The coefficients of space and time derivatives in
(\ref{Wave}) depend on background quantities, and for $\beta<0$ they have the
same sign for all values of inhomogeneity. Secondly, the same form of
signature change appears in spherically symmetric models in which
inhomogeneity need not be treated perturbatively.

\section{Related effects}
\label{s:Effects}

In the explicit form as provided by models of loop quantum gravity, signature
change is a new effect. But it has precursors in physics as well as
mathematics. Besides the examples discussed in details below, the case of
helically symmetric binary systems \cite{Helical} is worth mentioning. The
configuration is described by partial differential equations which change
signature along a spacelike direction (far from the center), rather than a
timelike one as in our cosmological models.

\subsection{Hartle--Hawking wave function}

A Lorentzian space-time cannot be closed off at a finite time without
a boundary. As proposed in \cite{nobound}, however, one may postulate that
quantum gravity gives rise to a modified space-time structure that allows a
transition to Euclidean 4-dimensional space. A Euclidean cap can then be
attached to a space-time manifold which becomes Lorentzian at low
curvature. In \cite{nobound} and the literature based on it, this scenario has
been used mainly to specify the high-density asymptotics of wave functions
satisfying the Wheeler--DeWitt equation of homogeneous models. 

The scenario suggested by effective constraints in models of loop quantum
gravity can be seen as a concrete realization of the quantum-gravity effects
that may give rise to signature change. The condition that
extrinsic curvature vanish at the interface between Euclidean and Lorentzian
parts of semiclassical solutions \cite{RealTunneling} is then replaced by
$\beta(h,K)=0$. Nevertheless, it is not guaranteed that the same
consequences are implied for wave functions, not the least because the
possibility of signature change relies on a quantum-geometry effect (based on
the use of holonomies in loop quantum gravity) which turns the Wheeler--DeWitt
equation into a difference equation \cite{cosmoIV,IsoCosmo}. The
minisuperspace discreteness implied by this difference equation is relevant
especially at high density, that is in the Euclidean phase where asymptotic
properties are discussed according to \cite{nobound}. Qualitatively different
conclusions for wave functions could then be reached, so that it is not clear
whether the close relationship in the space-time picture of \cite{nobound}
with the models discussed here implies a similar relationship in
predictions. Specific details of wave functions might well be closer to
\cite{tunneling} than to \cite{nobound}. We leave this question open in the
present article, in which we are concerned mainly with effective equations.

\subsection{Analog condensed-matter models}
\label{s:Analog}

In addition to related cosmological models in terms of space-time properties,
there are rather different physical phenomena which give rise to mathematical
descriptions similar to (\ref{Wave}). The main examples known to us are
transonic flow and hyperbolic metamaterials \cite{HypMM}.  Especially the
latter phenomenon provides an interesting analog picture of the cosmological
effects.

An example of a hyperbolic metamaterial is given by an array of parallel
conducting nano-wires immersed in a dielectric medium. Such a material behaves
as a conductor in the direction parallel to the nano-wires while in the normal
one it exhibits dielectric properties \cite{HypMmSign}. Choosing the $z$-axis
to be parallel to the nano-wires, the wave equation for the component $E^z =:
\phi$ of the electromagnetic field can be written as \cite{HypMmSign}
\begin{equation}
\frac{\partial^2 \phi}{\partial t^2}-\frac{1}{\epsilon_{\bot}}
\frac{\partial^2 \phi}{\partial z^2} 
-\frac{1}{\epsilon_{||}}\left(\frac{\partial^2 \phi}{\partial
    x^2}+\frac{\partial^2 \phi}{\partial y^2}\right)=0\,. 
\label{metamaterialEq}
\end{equation}
For the ``wired'' metamaterials the electric permittivity is
$\epsilon_{||}>0$, as for the standard dielectric medium. However, because the
metamaterial is conducting in the $z$-direction, $\epsilon_{\bot}$ will be
negative in a sufficiently low frequency range.\footnote{We assume here that
  the magnetic properties of the metamaterial are as usual, that is magnetic
  permeabilities in all directions are positive. However, it is worth
  mentioning that metamaterials with both electric permittivity $\epsilon<0$
  and magnetic permeability $\mu<0$ in some frequency band have been
  constructed. An interesting property of such metamaterials is that the
  refractive index is negative $n=-\sqrt{\epsilon \mu}$, leading to very
  interesting and sometimes counterintuitive behavior \cite{negref}.  The
  relevance of this phenomenon in the context of analog models of quantum
  gravity is an open issue.}  In particular, for the Drude theory of a
conducting medium
\begin{equation}
\epsilon_{\bot}=1-\frac{\omega^2_{\rm P}}{\omega^2}\,,
\end{equation}
where $\omega_{\rm P}$ is the plasma frequency dependent on the geometry and
composition of the metamaterial. At low frequencies ($\omega<\omega_{\rm P}$),
$\epsilon_{\bot}>0$ and the spatial part of equation (\ref{metamaterialEq}) is
elliptic. However, for $\omega>\omega_{\rm P}$, we have $\epsilon_{\bot}<0$
and the sign in front of the second derivative with respect to $z$
changes. The spatial Laplace operator becomes hyperbolic, with the
$z$-component playing a role of the second time variable. The dispersion
relation
\begin{equation}
\omega^2- \frac{k_z^2}{\epsilon_{\bot}}-\frac{k^2_{||}}{\epsilon_{||}}=0\,,
\end{equation}
where $k^2_{||}:=k_x^2+k_y^2$, for fixed $\omega$ is in this regime no longer
represented by ellipsoids but hyperboloids. In consequence, the $k$-space
undergoes a topology change.

Passing between the elliptic and hyperbolic regions is not only a matter of
frequency dependence.  For a fixed (sufficiently low) frequency, the transition
between the different signs of $\epsilon_{\bot}$ can be induced by a
rearrangement in the structural form of the ``wired" metamaterial.  In
particular, melting of the nano-wires to the liquid phase (associated
with a first-order phase transition) has been observed
experimentally. During the process the value of $\epsilon_{\bot}$ smoothly
changes its sign (from negative to positive) with increasing temperature.

Alternatively one can imagine the transition to be of the second order. This
idea in the context of signature change due to holonomy corrections has been
discussed in \cite{SigChange,Silence,Critical}. Let us imagine that the
nanowires are immersed in a dielectric fluid, having an ability to
rotate. Furthermore, let us introduce magnetic dipole-dipole type interactions
between the wires by attributing magnetic moments to them. Then, in the high
temperature unordered phase, with all nano-wires pointing in random
directions, there is hardly any electric conductivity. But when the phase
changes to an ordered one (lowering the temperature), a significant electric
current starts flowing, just as time starts flowing in our universe models
when the density is small enough to trigger signature change to Lorentzian
space-time.
 
In the gravitational case, the process can be modelled by a bi-metric gravity
theory, where the effective metric experienced by the fields is
\begin{equation}
g_{\mu\nu} = \delta_{\mu\nu}-2 \chi_{\mu} \chi_{\nu}\,.
\end{equation}
The metric may be viewed as a 4-dimensional analog of the inverse of the
electric permittivity tensor $\epsilon_{ij}$.  The $\chi_{\mu}$ is an
effective mean field having the interpretation of an order parameter, defined
such that in the unordered phase $|\vec{\chi}|=0$ (here $|\vec{\chi}| =
\sqrt{\delta^{\mu\nu} \chi_{\mu} \chi_{\nu}}$).  In this case the fields
experience the Euclidean metric $g_{\mu\nu} = \delta_{\mu\nu}$. In the fully
ordered phase $|\vec{\chi}|=1$, the Lorentzian signature emerges in a
spontaneously chosen direction in the 4-dimensional space.

In order to describe the transition from the unordered to the ordered state
more quantitatively let us postulate the following form of the free energy for
a model with a test field $\phi$:
\begin{eqnarray}
F = \int {\rm d}V  \underbrace{ \left(  \delta^{\mu\nu}+\frac{2 \chi^{\mu}
      \chi^{\nu}}{1-2 |\vec{\chi}|^2}  \right)}_{g^{\mu\nu}} 
\partial_{\mu} \phi \partial_{\nu} \phi +\underbrace{C \left[  \left(
      \frac{\rho}{\rho_{\rm QG}}-1\right)|\vec{\chi}|^2 
+\frac{1}{2}|\vec{\chi}|^4 \right]}_{V(\chi^{\mu},\rho)}\,,  \nonumber
\end{eqnarray} 
where $C$ is a constant. Because of the $\chi^{\mu} \chi^{\nu} \partial_{\mu}
\phi \partial_{\nu} \phi$ contribution to the kinetic term, $F$ is not
SO(4)-invariant in the internal indices of $\chi^{\mu}$. The explicit
symmetry-breaking term can be treated perturbatively. Then, equilibrium
corresponds to a minimum of the SO(4)-invariant potential
$V(\chi^{\mu},\rho)$. In order to model signature change in loop
  quantum cosmology with maximum energy density $\rho_{\rm QG}$, the
parameters of the potential are fixed such that at energy densities $\rho
> \rho_{\rm QG}$ the vacuum state maintains the SO(4) symmetry --- the order
parameter is $|\vec{\chi}|=0$.  For $\rho \leq \rho_{\rm QG}$, 
  we choose the potential so that its minimum is located at
$|\vec{\chi}|=\sqrt{1-\rho/\rho_{\rm QG}}$, with some spontaneously
chosen direction in the 4-dimensional configuration space of the order
parameter. Without loss of generality, we may assume
  $\vec{\chi}$ to point in the $0$-direction ($\chi_0=|\vec{\chi}|$ and
  $\chi_i=0$ for $i=1,2,3$); thus, $g_{00} =
1-2\chi_0\chi_0=-1+2\rho/\rho_{\rm QG} =-\beta$ and $g_{ii} = 1$. The
corresponding wave equation for the test field is
\begin{equation}
-\frac{1}{\beta}\frac{\partial^2}{\partial t^2} \phi+\Delta \phi= 0\,,  
\label{EQMv}
\end{equation}
which is of the form (\ref{Wave}) with $\beta=1-2\rho/\rho_{\rm QG}$ as in 
models of loop quantum cosmology. At $|\vec{\chi}|=0$ ($\beta=1$) the 
equation exhibits the $R^4 \rtimes {\rm SO}(4)$ symmetry while in the fully 
ordered state $|\vec{\chi}|=1$ ($\beta=-1$) the $R^{1,3} \rtimes {\rm SO}(1,3)$ 
Poincar\'e symmetry is satisfied. 

It is worth stressing that the presented model relating signature change with
the spontaneous symmetry braking of $SO(4)$ has no support from more
fundamental considerations at the moment.  Therefore, its physical validity
requires further studies. In particular, a possible relation of the order
parameter $\chi_{\mu}$ with the elementary degrees of freedom of loop quantum
gravity remains unclear. Nevertheless, the model presents an interesting
example because of its formal resemblance to signature change in loop quantum
cosmology. There is, however, an important conceptual difference: while analog
models of signature change are fully well-defined and even have some
observable properties, quantum-cosmology effects on space-time itself (rather
than matter fields in space-time) may be subject to several global problems on
which we will comment in Sec.~\ref{s:Global}.

\subsection{Mixed-type characteristic problem and Tricomi equation}

Partial differential equations of mixed type have been studied in the
mathematics literature since the 1930s, beginning with the Tricomi problem for
the differential equation
\begin{equation}
\frac{\partial^2 u}{\partial y^2}+y\frac{\partial^2u}{\partial x^2}=0
\label{TricomiEQ}
\end{equation}
in the $(x,y)$-plane. This type of equation can be seen as an approximation to
our equation (\ref{Wave}) around one of the finite boundaries of the elliptic
regime, where $\beta=0$. Here, we recall relevant features of a suitable
characteristic problem \cite{Tricomi,TricomiGen}.

\begin{figure}
\begin{center}
\includegraphics[width=12cm]{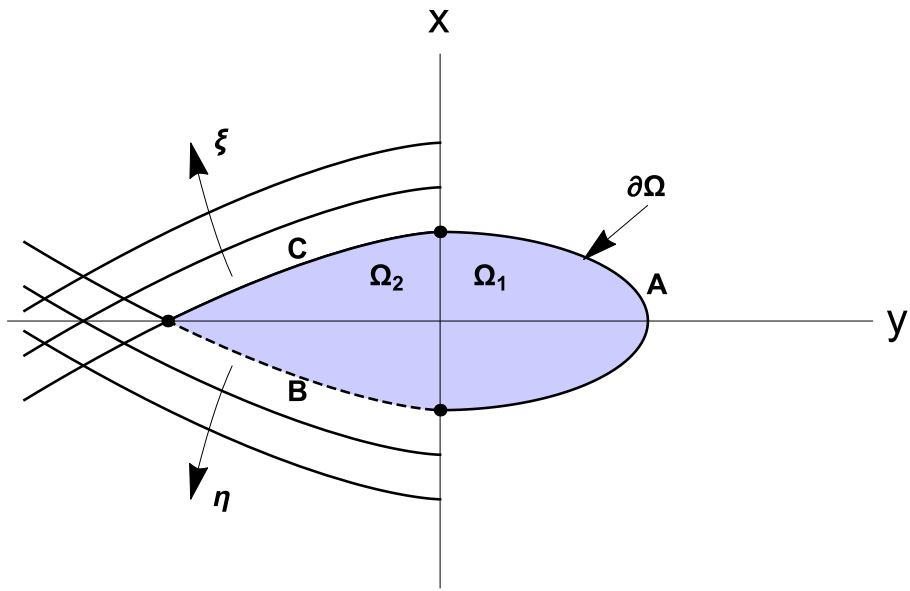}
\caption{Graphical representation of the characteristic problem from the
  mixed-type Tricomi equation. Solutions to a partial
    differential equation (\ref{TricomiEQ}) (hyperbolic for $y<0$ and elliptic
    for $y>0$) exist uniquely and stably in the shaded region
    $\Omega_1\cup\Omega_2$, provided boundary data are imposed on one
    characteristic $C$ of the hyperbolic region together with a curve $A$ in
    the elliptic region. Characteristic coordinates (\ref{charac}) are
    indicated by $\xi$ and $\eta$.}
\label{Tricomi1}
\end{center}
\end{figure}

Let us suppose that we would like to solve equation (\ref{TricomiEQ}) in a
domain $\Omega=\Omega_1\cup \Omega_2$, where $\Omega_1$ and $\Omega_2$ are
elliptic and hyperbolic domains, respectively.  For the Tricomi problem, the
sub-domain $\Omega_2$ cannot be chosen in an arbitrary manner.  Namely, the
boundary $\partial \Omega_2$ is determined by characteristics of
(\ref{TricomiEQ}) attached to the endpoints of $\partial \Omega_1$ at the
interface between the hyperbolic and elliptic regions; see
Fig.~\ref{Tricomi1}. The characteristics in the hyperbolic region are
determined by the characteristic equation to (\ref{TricomiEQ}): $({\rm
  d}x)^2+y({\rm d}y)^2=0$. This equation has two families of solutions with a
different sign, enabling us to introduce the characteristic coordinates
\begin{eqnarray}
  \xi := x+\frac{2}{3}(-y)^{3/2}\,,  \\
  \eta := x-\frac{2}{3}(-y)^{3/2}\,. 
\end{eqnarray} 
(We have $y<0$ in the hyperbolic region.)

Solutions to the Tricomi equation (\ref{TricomiEQ}) are unique and stable if
adequate boundary conditions are imposed. Applying the notation from
Fig.~\ref{Tricomi1}, they are as follows: in region $\Omega_1$, the value of
$u$ has to be specified at the boundary curve $A$, while in region $\Omega_2$
the value of $u$ at the characteristic curve $C$ has to be given.  The value
of $u|_{B}$ is a result of fixing the boundary condition $u|_{C\cup A}$. 

The reason why the boundary condition in the hyperbolic region is imposed on
the characteristic curve becomes clear by analyzing the case of the standard
wave equation with one spatial dimension:
\begin{equation}
\frac{\partial^2 u}{\partial y^2}-\frac{\partial^2u}{\partial x^2}=0\,.
\label{WaveEQ}
\end{equation}
The corresponding characteristic equation $({\rm d}x)^2-({\rm d}y)^2=0$ has
solutions $x\pm y=C$, which allow us to introduce the characteristic (light
cone) variables
\begin{eqnarray} \label{charac}
\xi := x+y\,,  \\
\eta := x-y\,. 
\end{eqnarray} 
In terms of these variables the wave equation (\ref{WaveEQ}) reduces to the
canonical form
\begin{equation}
\frac{\partial^2 u}{\partial \xi \partial \eta}=0\,,
\end{equation}
with general solution $u=f(\xi)+g(\eta)$. An important observation is that an
initial-value problem (Cauchy problem) in the variables $(x,y)$ translates
into the boundary-value problem in the characteristic variables
$(\xi,\eta)$. In particular, the Cauchy-problem initial condition $u|_{y=0}$
and $\left. \partial u/\partial y \right|_{y=0} $ translates into the
Dirichlet boundary condition at the light cone $u|_{\xi=0}$ and $u|_{\eta=0}$.

Using characteristics, it is therefore easier to combine hyperbolic and
elliptic regimes. In the case of the Tricomi equation, the boundary conditions
in the elliptic domain have to be smoothly extended into the hyperbolic
sector. This requires the introduction of characteristic variables in the
hyperbolic sector, imposing boundary conditions at either $\xi=$ constant or
$\eta=$ constant (but not both).

\section{Cosmology}
\label{s:Cosmo}

Signature change plays an important role in cosmological scenarios. In
practical terms, one of the most important consequences is the presence of
instabilities, as indicated formally by an imaginary speed of sound in mode
equations. The same kind of instability plays a role in the mathematical
discussion of equations of the form (\ref{Wave}): One of the conditions for a
well-posed initial/boundary-value problem of a partial differential equation
is the stable dependence of solutions on the chosen data, in addition to the
conditions that a solution exist and be unique for given data.

\subsection{Instability}

Stability is easily violated when one attempts to use an initial-value problem
for an elliptic equation. For instance, Fourier modes of solutions to the
standard wave equation would not oscillate as per $\exp(\pm i\omega t)$ but
change exponentially according to $\exp(\pm\omega t)$. The exponentially
growing mode implies instability in the sense that solutions, if they exist
and are perhaps unique, depend sensitively on the initial values. Choosing a
boundary-value problem in $t$ (as well as spatial directions), on the other
hand, ties down the solution at both ends of the $t$-range, so that the
growing branch is sufficiently restricted for solutions to be stable.

Signature change implies instability of initial-value problems, but it is
stronger in two respects. First, signature change affects all modes (of matter
or gravity) equally, which all become unstable at the same ``time.'' It is
therefore a space-time effect, rather than an exotic matter
phenomenon. Secondly, it can be seen in space-time symmetries such as
hypersurface deformations or the Poincar\'e algebra. None of these effects
happen in known cases of instabilities in matter theories or higher-curvature
gravitational actions, such as \cite{CdD}. Signature change is therefore more
fundamental than other phenomena hat might imply instability. An important
property is the fact that the theory does not provide any causal structure
whatsoever when the signature turns Euclidean.

\subsection{Scenarios}

When used in cosmological model-building, signature change implies a new and
interesting mixture of linear and cyclic models. One of its main consequences
can be described as a finite beginning of the universe. In the simplest
versions of existing models of loop quantum cosmology, effective equations do
not imply divergences at high density but instead extend solutions to a new
low-density regime \cite{QuantumBigBang}. The main example is a modified
Friedmann equation
\begin{equation} \label{ModFried} 
\frac{\sin(\delta \bar{\cal
      H})^2}{\delta^2}= \frac{8\pi G}{3}\rho
\end{equation}
postulated for spatially flat isotropic models sourced by a free, massless
scalar $\phi$. The energy density $\rho$ is therefore of the form
$\rho=\frac{1}{2} p_{\phi}^2/a^6$ with the momentum $p_{\phi}$ of $\phi$ and
the scale factor $a$. In (\ref{ModFried}), $\delta$ is a parameter which
approaches $\delta\to0$ in the classical limit, and whose precise value
remains undetermined. (As a parameter motivated by quantum-gravity effects, it
is often assumed to be close to the Planck length.) Moreover, $\bar{\cal
  H}=(2\delta)^{-1} \arcsin(2\delta\dot{a}/a)$ is a modified version of the
Hubble parameter. It is straightforward to see that the modified equation
(\ref{ModFried}) implies that the energy density, for fixed $\delta$, is
always bounded, $\rho \leq \rho_{\rm QG} := 3/(8\pi G \delta^2)$.  The
parameter $\rho_{\rm QG}$ is the maximum kinetic energy density realized at
the bounce. There is then a good chance that cosmological singularities may be
resolved. Indeed, the quantum model with a free, massless scalar can be solved
completely \cite{BouncePert}, implying that the scale factor
\begin{equation} \label{at}
 a(t)= a_0\left(1+ 24\pi G\rho_{\rm QG}\: t^2\right)^{1/6}
\end{equation}
never becomes zero. (See also \cite{BounceSols}.)

Equation (\ref{ModFried}), although it remains unclear how generically it is
realized for the background evolution \cite{BounceSqueezed} of general
homogeneous models, has suggested a picture in which our expanding universe
descends from a preceding collapse phase which produced large but not infinite
density at the big bang. However, when modes on such a bouncing background are
subject to equations of the form (\ref{Wave}), as required for a consistent
system of both background variables $(a,\phi)$ and inhomogeneous modes in
models of loop quantum gravity, the transition is not deterministic owing to
the lack of causal structure at high density. In the cosmological case, the
role of $K$ in (\ref{Wave}) is played by the modified Hubble parameter
$\bar{\cal H}$. With $\beta(\bar{\cal H})\propto \cos(2\delta\bar{\cal H})$,
one can easily see that signature change happens when the energy density is
half the maximum it can achieve (or half the bounce density of a homogeneous
model). Given the solution (\ref{at}), the mode equation (\ref{Wave}) is
elliptic for
\begin{equation} \label{t}
 t^2 < t_{\rm max}^2:=\frac{1}{24\pi G \rho_{\rm QG}}\,.
\end{equation}
(For $\rho_{\rm QG}$ close to the Planck density, the $t$-range in which the
equation is elliptic is close to the Planck time, but it can be significantly
longer in models with a scalar potential or strong quantum fluctuations at
high density \cite{FluctEn}.)  The high-density regime cannot be accessed
causally, and there is no deterministic transition from collapse to expansion.
For practical purposes, such a scenario therefore shares some features with
the traditional singular big-bang model, in which one must pose initial values
just after the singularity. Similarly, one must pose initial values ``after''
the Euclidean phase of a signature-change scenario. However, the relationship
between singular and signature-change models is subtle, as shown by the
detailed discussion of suitable data for mixed-type differential equations on
which we now embark.

\subsubsection{Gravitational waves}
\label{s:GW}

To be more specific, let us focus now on the case of gravitational waves with
holonomy corrections. In this case, the equation of motion for each component
of polarization of the gravitational waves is \cite{ScalarTensorHol}
\begin{equation}
\frac{\partial^2 \phi}{\partial
  t^2}+\left(3H-\frac{\dot{\beta}}{\beta}\right)\frac{\partial \phi}{\partial
  t}-  \frac{\beta}{a^2}\Delta \phi =0\,, 
\label{EOMphi}
\end{equation}
where $H :=\frac{\dot{a}}{a}$ is the Hubble parameter and the deformation
factor is
\begin{equation} \label{betarho}
\beta = \cos(2\delta \bar{\cal H}) = 1-2\frac{\rho}{\rho_{\rm QG}}\,.  
\end{equation}
Depending on the value of the parameter $\beta$, equation (\ref{EOMphi})
can be either hyperbolic ($\beta > 0$), elliptic ($\beta < 0$) or parabolic
($\beta = 0$). The transition between elliptic and hyperbolic type (associated
with signature change) takes place at $\rho = \frac{1}{2} \rho_{\rm
  QG}$. At this moment, the square of the Hubble factor $H$ reaches its maximal
value $H^2_{\rm max} = \frac{2}{3}\pi G \rho_{\rm QG}$. For the model with a
free scalar field introduced in the previous section, this takes place at $t =
\pm t_{\rm max} = \pm (24\pi G \rho_{\rm QG})^{-1/2}$.

Equation (\ref{EOMphi}) is precisely of the form (\ref{Wave}) with the source
term
\begin{equation}
S[\phi]=-\left(3H-\frac{\dot{\beta}}{\beta}\right)\frac{\partial
  \phi}{\partial t}\, .
\end{equation}
Because generally $\dot{\beta}\neq 0$ at $\beta=0$, signature change is
associated with a divergence of the source. The divergence, however, does not
lead to pathological behavior at the level of solutions to equation
(\ref{EOMphi}) because, as we will see later, it is due to a regular pole,
which does not disturb regularity of the solution.  In the vicinity of
$\beta=0$, equation (\ref{EOMphi}) can be approximated by
\begin{equation}
\frac{\partial^2 \phi}{\partial
  t^2}+\left(3H-\frac{\dot{\beta}}{\beta}\right)\frac{\partial \phi}{\partial
  t}\approx0\, , 
\end{equation}
which after double integration leads to the solution
\begin{equation}
\phi = c_1+c_2 \int^t \frac{\beta(t')}{a(t')^3}dt', 
\label{ApproxSol1}
\end{equation}
where $c_1$ and $c_2$ are some functions of the spatial variables. Because
$\beta$ occurs in the numerator of the integrand, the approximate solution is 
indeed regular.

Let us now analyze the behavior of equation (\ref{EOMphi}) in the vicinity of
the instance of signature change for the model with a free scalar
field. Without loss of generality, we perform an expansion around $t=-t_{\rm
  max}$, for which
\begin{equation}
\left(3H-\frac{\dot{\beta}}{\beta}\right) = \frac{t (t^2-5t^2_{\rm
    max})}{t^4-t^4_{\rm max}} = 
-\frac{1}{t+t_{\rm max}}-\frac{1}{t_{\rm max}}+\frac{t+t_{\rm max}}{4t^2_{\rm
    max}}+\mathcal{O}\left((t+t_{\rm max})^2\right)\, .
\end{equation}
(In the first step we have used (\ref{at}), (\ref{t}) and (\ref{betarho}) in
order to write $a(t)=a_0(1+t^2/t_{\rm max}^2)^{1/6}$ and
$\beta(t)=1-2(1+t^2/t_{\rm max}^2)^{-1}$.)  Because $(t+t_{\rm
  max})\left(3H-\dot{\beta}/\beta\right)=-1+ \mathcal{O}\left(t+t_{\rm
    max}\right)$, the pole at $t=-t_{\rm max}$ is of the regular
type. Furthermore
\begin{equation}
\frac{\beta}{a^2}= \frac{(t/t_{\rm max})^2-1}{(1+(t/t_{\rm max})^2)^{4/3} } 
= - \frac{t+t_{\rm max}}{2^{1/3}t_{\rm max}}  +\mathcal{O}\left((t+t_{\rm
    max})^2\right)\, ,
\end{equation}
where we fixed $a_{0}=1$. 

Applying the above expansions to the leading order and reducing to
the (1+1)-dimensional case\footnote{Of course this reduction 
is formal only because gravitational waves do not occur in (1+1)-dimensions. 
Alternatively, one can consider the reduction as being a result of 
introducing translational symmetry in the $y$ and $z$ directions, leading 
to $\phi(t,x,y,z) = \phi(t,x)$.}, we obtain
\begin{equation}
\frac{\partial^2 \phi}{\partial t^2}-\frac{1}{t+t_{\rm max}}\frac{\partial
  \phi}{\partial t} 
+\frac{t+t_{\rm max}}{2^{1/3}t_{\rm max}}\frac{\partial^2 \phi}{\partial x^2}=0\,. 
\label{ReducedEOM1}
\end{equation}
Redefining the time variable to 
\begin{equation}
y:= \frac{1}{\left(2^{1/3} t_{\rm max}\right)^{1/3}}  (t+t_{\rm max})\,,
\end{equation}
equation  (\ref{ReducedEOM1}) can be rewritten as
\begin{equation}
\frac{\partial^2 \phi}{\partial y^2}+y\frac{\partial^2\phi}{\partial
  x^2}=\frac{1}{y}  \frac{\partial \phi}{\partial y}\, . 
\label{ALaTricomi}
\end{equation}
The left-hand side of this equation is identical to Tricomi's expression in
(\ref{TricomiEQ}) while the right-hand side can be treated as a source
term. Applying the characteristic Tricomi problem to obtain solutions to
the equation (\ref{ALaTricomi}) is therefore justified. 

It is worth noting at this point that equation (\ref{ALaTricomi}) in the 
vicinity of $y=0$ reduces to
\begin{equation}
\frac{\partial^2 \phi}{\partial y^2}-\frac{1}{y}  \frac{\partial
  \phi}{\partial y}=0\, ,  
\label{ALaTricomiReduced}
\end{equation}
with solution $\phi = c_1+c_2y^2$ (in agreement with (\ref{ApproxSol1})).  The
solution is perfectly regular across the signature change, in contrast to
classical signature change. In particular, signature change resulting from the
line element ${\rm d}s^2 = -t{\rm d}t^2 +{\rm d}x_a{\rm d}x^a$ has been widely
studied in the literature \cite{ClassSigChange,ClassSigChange2,
  SigChangeJunction}. In this case, the analog of equation
(\ref{ALaTricomiReduced}) is $\frac{\partial^2 \phi}{\partial
  y^2}-\frac{1}{2y} \frac{\partial \phi}{\partial y}=0$, with the metric
signature change at $y=0$. While both forms of the equation look very similar,
the extra factor of ``2'' plays a significant role leading to the
non-differentiability of solution at the signature change. For discussion of
controversies around this issue as well as some proposals of dealing with the
problem we refer to Ref. \cite{SigChangeJunction}.

Coming back to expression (\ref{ALaTricomi}), the corresponding equation for 
the Fourier component $\phi_k = (2\pi)^{-1/2}\int {\rm d}ke^{-ikx}\phi(x)$ is
\begin{equation}
\frac{\partial^2 \phi_k}{\partial z^2}-\frac{1}{z}\frac{\partial
  \phi_k}{\partial z}-z\phi_k=0\, ,
\label{EOMphik}
\end{equation}
where $z:=y|k|^{2/3}$ has been introduced to absorb the wave number $k$.  In
order to facilitate solving this equation, we observe that by differentiating
the Airy equation
\begin{equation}
\frac{{\rm d}^2u}{{\rm d}z^2}-zu=0\,,
\label{Airy}
\end{equation}
and subsequently substituting for $u$ using again (\ref{Airy}), we find an
equation
\begin{equation}
\frac{{\rm d}^3u}{{\rm d}z^3}-\frac{1}{z}\frac{{\rm d}^2u}{{\rm d}z^2}
-z\frac{{\rm d}u}{{\rm d}z}=0
\end{equation} 
identical to (\ref{EOMphik}) if $\phi_k={\rm d}u/{\rm d}z$ for a suitable
($k$-dependent) $u$ solving (\ref{Airy}).  Solutions to (\ref{EOMphik}) are
therefore
\begin{equation}
\phi_k = A_k {\rm Ai}'(z)+B_k {\rm Bi}'(z)\,, 
\end{equation}
where ${\rm Ai}'(z)$ and $ {\rm Bi}'(z)$ are the Airy prime functions
(derivatives of the Airy functions) and $A_k$ and $B_k$ are $k$-dependent
constants of integration. Because of the reality condition for the field
$\phi$ the following relations have to be fulfilled: $A^*_k=A_{-k}$ and
$B^*_k=B_{-k}$.

\begin{figure}
\begin{center}
\includegraphics[width=12cm]{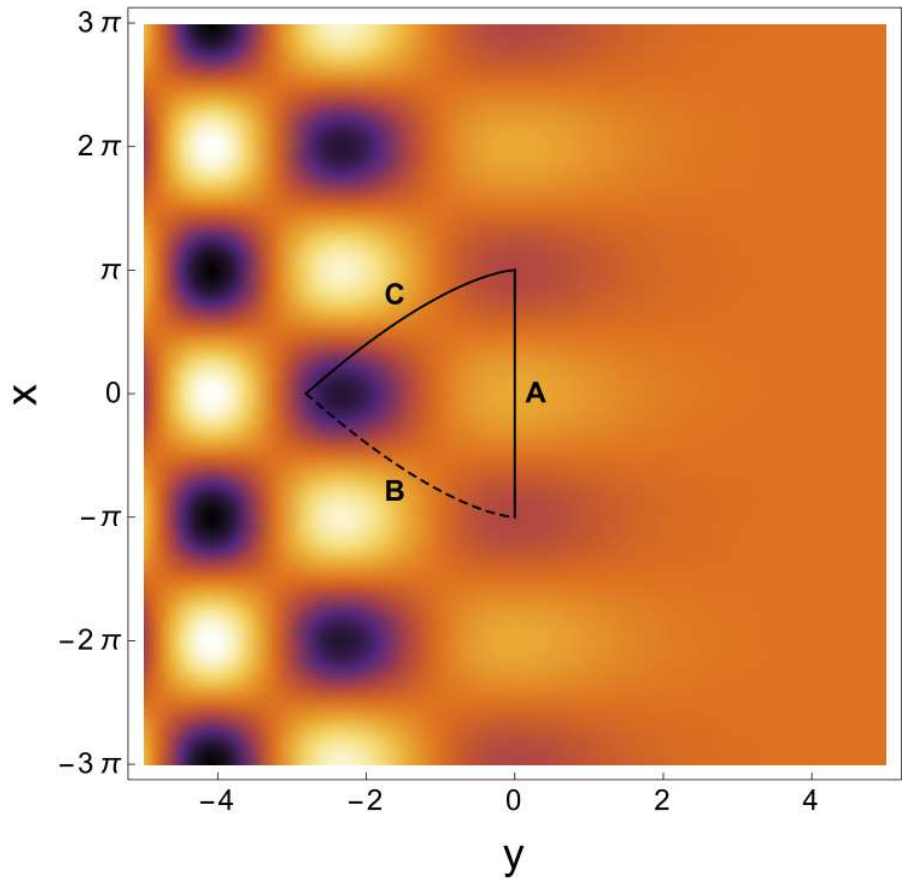}
\caption{Density plot of the solution (\ref{AlaTricomiSolutionFinal}) together
  with the boundary $\partial \Omega = A\cup B\cup C$. This special solution
  is exponentially decreasing for $y>0$, as can be seen by the fading-away of
  oscillations.}
\label{Solution}
\end{center}
\end{figure}

As an example, we will now apply the Tricomi boundary conditions to the
resulting 2-dimensional solution
\begin{equation}
\phi(y,x) = \int_{-\infty}^{+\infty} \frac{{\rm d}k}{\sqrt{2\pi}}
e^{ikx}\left[A_k {\rm Ai}'\left(y|k|^{2/3} \right)+B_k {\rm
    Bi}'\left(y|k|^{2/3} \right)\right]\, .
\label{AlaTricomiSolution}
\end{equation}
First, let us choose the contour $A := \left\{ (x,y) \in \mathbb{R}^2, y=0 ,
  -\pi<x<\pi \right\}$ and fix $\phi|_{A}= \cos{x}$.  This condition reduces
(\ref{AlaTricomiSolution}) to
\begin{equation}
\phi(y,x) =  \cos(x) \left[   \frac{1-c_0 {\rm Bi}'(0)}{ {\rm Ai}'(0)}  {\rm
    Ai}'(y)+c_0  {\rm Bi}'(y) \right]\,, 
\label{AlaTricomiSolutionReduced1}
\end{equation}      
with a parameter $c_0 \in \mathbb{R}$. Secondly, the form of $A$ implies
that:
\begin{eqnarray}
C&:=& \left\{ (x,y) \in \mathbb{R}^2,  x=\pi-\frac{2}{3}(-y)^{3/2}, -\left(
    \frac{3\pi}{2}\right)^{2/3}<y<0 \right\}\, , \\
B&:=& \left\{ (x,y) \in \mathbb{R}^2,  x=-\pi+\frac{2}{3}(-y)^{3/2}, -\left(
    \frac{3\pi}{2}\right)^{2/3}<y<0 \right\}\, .
\end{eqnarray}
Finally, by choosing 
\begin{equation}
\phi|_{C} = \cos\left(\pi-\frac{2}{3}(-y)^{3/2}\right) {\rm Ai}'(y)/{\rm Ai}'(0)
\end{equation} 
the constant  $c_0$ is fixed to equal zero, which leads to 
\begin{equation}
\phi(y,x) = - 3^{1/3} \Gamma(1/3) \cos(x){\rm Ai}'(y)\,. 
\label{AlaTricomiSolutionFinal}
\end{equation} 
The solution together with the boundary $\partial \Omega = A\cup B\cup C$ is
shown in Fig.~\ref{Solution}. For all other values of $c_0$, the contribution
from ${\rm Bi}'(y)$ in (\ref{AlaTricomiSolutionReduced1}) does not vanish, and
therefore the solution is characterized by an exponential growth in the $y>0$
region. Nevertheless, in the domain enclosed by $\partial \Omega = A\cup B\cup
C$ the solution remains regular. It is transparent from Fig.~\ref{Solution}
that the monotonic solution in the elliptic region ($y>0$) transforms smoothly
into the oscillatory solution in the hyperbolic domain ($y<0$), as expected
intuitively.

\subsubsection{Characteristic problem}
\label{s:Char}

We are now in a position to adapt the Tricomi problem to
effective space-time models of loop quantum gravity. We 
begin with a qualitative discussion which also provides 
heuristic reasons for why the mixed-type initial-boundary 
value problem must be used.

In a non-singular universe model in which collapse is joined by expansion, one
can start with initial values in the collapsing Lorentzian phase in the past
and evolve all the way to the boundary with the Euclidean phase (at $t=-t_{\rm
  max}$ in the solvable model). For a second-order hyperbolic differential
equation in this regime, we need to fix initial values $\phi$ and $\phi'$,
where we denote by $\phi'$ the derivative of $\phi$ normal to constant-$t$
hypersurfaces. In this regime, $\phi'$ may therefore be considered as the time
derivative of $\phi$. When we reach the Euclidean phase, time no longer exists
in a causal sense and our coordinate $t$ is one of the four spatial ones. We
will continue denoting the $t$-derivative as $\phi'$, now strictly in the
sense of a normal derivative replacing the time derivative.

\begin{figure}
\begin{center}
\includegraphics[width=12cm]{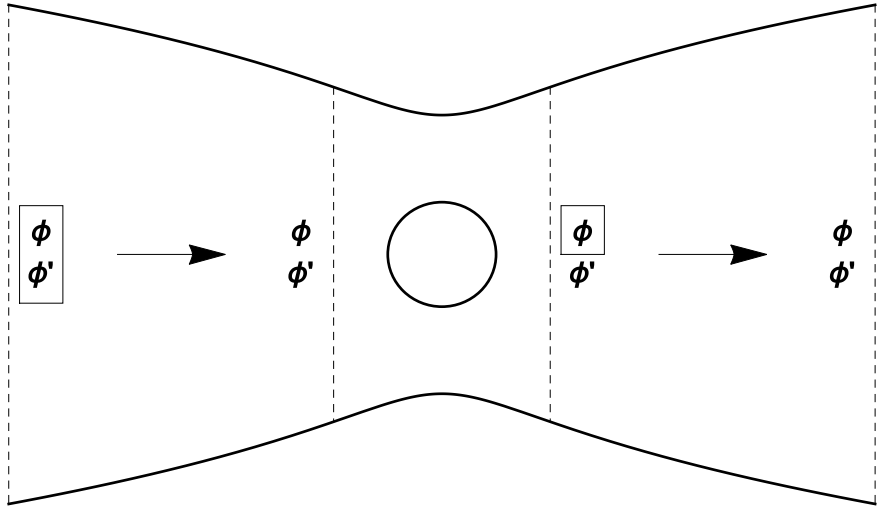}
\caption{Illustration of indeterministic behavior implied by the mixed-type
  differential equations of signature-change cosmology. The boxed data for a
  field $\phi$ --- the field and its first normal derivative $\phi'$ in the
  past Lorentzian phase as well as the field $\phi$ (but not its normal
  derivative) at the beginning of the expanding phase --- must be
  prescribed. The other data then follow from a solution to the mixed-type
  differential equation: $\phi$ and $\phi'$ at the left boundary of the
  Euclidean region by an initial-value problem in the past Lorentzian phase
  (left arrow); $\phi'$ at the right boundary of the Euclidean phase as the
  limiting normal derivative of the solution in the Euclidean phase based on
  the evolved $\phi$ on the left boundary and the prescribed boundary data
  $\phi$ on the right boundary (center circle); and finally the future field
  $\phi$ and $\phi'$ in the expanding phase evolved from initial values at the
  right boundary of the Euclidean phase (right arrow). This
  initial/boundary-value formulation is made more precise in the Tricomi
  problem, shown in Fig.~\ref{Fig:Tricomi}. \label{Fig:IB}}
\end{center}
\end{figure}

Once we reach the Euclidean phase and enter the range between $t=-t_{\rm max}$
and $t=t_{\rm max}$ in the model, we must switch to a boundary-value problem
for well-posedness. We can use our field $\phi$ evolved up to the part of the
Euclidean boundary bordering the collapse phase, at $t=t_{\rm max}$. But we
must complete it to a set of boundary values on a hypersurface enclosing the
Euclidean region in which we are looking for a solution. For a solution in all
of space(-time), this complete boundary includes asymptotic infinity in
directions other than $t$ as well as the boundary at $t=t_{\rm max}$ where the
Euclidean region borders on the expanding phase. We may take for granted that
asymptotic fall-off conditions for our fields are imposed at spatial infinity,
but we must choose a new and arbitrary function on the border with the
expanding phase. When this function is fixed, together with the evolved past
initial data, we obtain a solution in the Euclidean phase. We may then compute
normal derivatives on all constant-$t$ hypersurfaces in this region and take
the limit toward the border with the expanding phase. We obtain suitable data
for an initial-value problem in the latter. In summary, as illustrated in
Fig.~\ref{Fig:IB}, in addition to past initial data we must specify one
additional function for every mode at the beginning of the expansion
phase. Past initial data in the collapse phase do not determine a unique
solution in the expansion phase. There is no deterministic evolution across
the Euclidean high-density regime.

The intuitive description just given provides the correct picture of free and
determined data, but it is not the most suitable one for the problem. One
question one may pose is about the smoothness of solutions: A solution to
(\ref{Wave}) should be required to be smooth (or at least twice
differentiable) at the borders between Lorentzian and Euclidean regions. In
the picture of Fig.~\ref{Fig:IB}, it is not clear whether this requirement is
satisfied. We have to match two solutions, one in the past Lorentzian phase
and the one in the Euclidean phase, so that the normal derivative $\phi'$
obtained from these solutions changes differentiably across the border. Since
the field $\phi$ on the border with the expanding phase is free and affects
the solution in the Euclidean region but not in the collapse phase,
differentiability seems in danger unless one severely restricts the boundary
data left free around the Euclidean region: If we keep past initial data fixed
but vary the boundary field $\phi$ at the border $t=t_{\rm max}$ with the
expansion phase, the limit of $\phi'$ at $t=-t_{\rm max}$ will change for
solutions of the elliptic equation. Only a narrow range of boundary fields
$\phi$ would seem to provide a $\phi'$ matching the evolved initial data
in the collapse phase. If this is so, one would essentially be led back to an
initial-value problem across the whole domain, because $\phi$ and $\phi'$
everywhere would be uniquely determined by initial values, a conclusion which
cannot be correct if there is an elliptic regime. By these considerations, for
a smooth solution we would have to give up some of the initial values in the
collapse phase as free data, but using just $\phi$ and letting $\phi'$ be
determined by smoothness cannot be correct either, because we would end up
with a boundary-value problem which is not well-posed in the Lorentzian phase.

\begin{figure}
\begin{center}
\includegraphics[width=12cm]{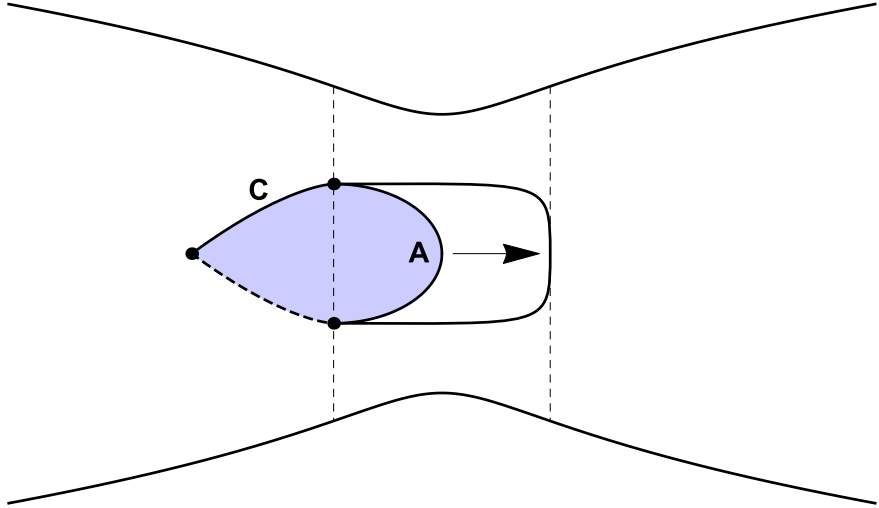}
\caption{Well-posed characteristic problem for a mixed-type partial
  differential equation: The function is specified on one characteristic arc
  $C$ of the equation in the Lorentzian phase and on an arc $A$ in the
  Euclidean phase connecting the endpoints of $C$ and of a second
  characteristic arc starting at the same point as $C$ (dashed). The data on
  the second characteristic arc are determined for solutions of (\ref{Wave})
  and cannot be chosen freely.  \label{Fig:Tricomi}}
\end{center}
\end{figure}

In order to address differentiability, a Tricomi-style characteristic problem
is more appropriate. In the Lorentzian phase, one specifies data not on
initial-value surfaces but on characteristics of the differential equation as
long as it is hyperbolic. Far away from the border with the Euclidean region,
these characteristics are standard light rays, but they approach the border
normal to constant-$t$ surfaces and end there. One example is shown as curve
$C$ in Fig.~\ref{Fig:Tricomi}. Considering just one spatial dimension, a
second characteristic starting from the same point as $C$ completes the first
one to a characteristic triangle with the two characteristics as well as the
enclosed border with the Euclidean region as its sides. 

The problem is well-posed \cite{Tricomi,TricomiGen} if one specifies the field
$\phi$ on one of the two characteristics and on an arc $A$ which starts and
ends at the border-endpoints of the two characteristics but stays in the
Euclidean region in between. If these data are smooth on the curve obtained as
the union of $C$ and $A$ one obtains a unique, stable and smooth solution in
the interior of the union of the characteristic triangle and the region
enclosed by the curve $A$. Smoothness is guaranteed in this way, with data on
the curve $A$ being completely free except that they must extend the data on
$C$ smoothly. The indeterministic behavior (requiring future data) is
therefore confirmed. If one lets $A$ approach the other border of the
Euclidean region, one can proceed as in the intuitive description based on
Fig.~\ref{Fig:IB}. An initial-value problem starting at the beginning of the
expanding phase is then well-posed toward the future.

In four dimensions, the characteristic triangle is extended to the interior of
the future light cone starting somewhere in the collapse phase, and the region
enclosed by a curve $A$ is extended to a dome enclosed by a 3-dimensional
hypersurface in the Euclidean region.  Although we are not aware of a complete
mathematical proof, one may expect the characteristic problem to be well-posed
with data given by the mode on half the light cone and on the hypersurface in
the Euclidean region.

The smoothness of solutions shows that there is still a well-defined manifold
which combines the two Lorentzian phases and the Euclidean phase, but this
manifold is no longer Riemannian. Standard obstacles to signature change in
general relativity therefore do not apply. The use of a finite region in the
characteristic formulation does not cause a problem in cosmology because we
always have observational access to a finite part of our universe. The
characteristic triangle can always be chosen large enough to include the whole
observational range.

\subsubsection{Boundary conditions at the hypersurface of signature change}

Besides the boundary, initial-value and the characteristic problems discussed
in the previous subsection we would like to present one more possibility of
posing the initial-value problem for the evolution in the Lorentzian
domain. Let us start by finding the solution in the Euclidean domain.

In this domain (which we call $\Omega_E$) the equations of motion are of the
elliptic type. Therefore, for finding solutions in a subset $ \Omega\subset
\Omega_E$ the proper boundary conditions have to be imposed.  Specifying $\phi
|_{\partial \Omega}$ (where $\partial \Omega$ is the boundary of $\Omega$) is
sufficient to determine a solution $\phi_E$ to the elliptic equation in the
domain $\Omega$ uniquely.  An interesting situation corresponds to the case
when the boundary $\partial \Omega$ is expanded so that it encloses the whole
Euclidean domain: $\partial \Omega \rightarrow \partial \Omega_E$.  Such a
boundary can be decomposed as follows $\partial \Omega_E = \partial
\Omega_{+}+ \partial \Omega_{-} + \partial \Omega_{\infty}$. The $\partial
\Omega_{\pm} $ are boundaries between the Euclidean and Lorentzian domains at
$\pm t_{\rm max}$, respectively, while $\partial \Omega_{\infty}$ encloses the
Euclidean domain at spatial infinity (assuming that the space part is
unbounded). The solution $\phi_E$ at $\Omega_E$ obtained by imposing boundary
(Dirichlet or von Neumann) conditions at $\partial \Omega_E$ can now be used
to determine initial (or final) conditions for the Lorentzian sectors
$\Omega_{\pm}$, where $``+"$ corresponds to $t>+t_{\rm max} $ and $``-"$ to
$t<-t_{\rm max}$.  The $\pm t_{\rm max}$ (as previously) are the times at
which signature change is taking place.

Let us decide to impose the Dirichlet boundary condition $\phi_E |_{\partial
  \Omega_E}$, leading to a solution $\phi_E$. Then, the Cauchy initial
conditions at $\pm t_{\rm max}$ for the solutions $\phi_{L\pm}$ in the
Lorentzian domains $\Omega_{L\pm}$ respectively are:
\begin{eqnarray}
\lim_{t\rightarrow \pm {t_{\rm max}}^{\pm}} \phi_{L\pm} &=& \phi_E|_{\partial
  \Omega_{\pm}} \equiv \phi_{\pm}, \\     
\lim_{t\rightarrow \pm {t_{\rm max}}^{\pm}}\frac{\partial \phi_L }{\partial t}  &=& 
\lim_{t\rightarrow \pm {t_{\rm max}}^{\mp} }  \frac{\partial \phi_E }{\partial
  t} \, .
\end{eqnarray}
In this way by solving the equation of motion in the whole Euclidean domain,
the initial (or final) data for evolution in the Lorentzian sector can be
recovered.  It is worth stressing that, as already discussed in the previous
subsection, the continuity of solutions up to at least second time derivatives
is required. (The equations are of second order.) The additional conditions
\begin{equation}
\lim_{t\rightarrow \pm {t_{\rm max}}^{\pm}}\frac{\partial^2 \phi_L }{\partial t^2} = 
\lim_{t\rightarrow \pm {t_{\rm max}}^{\mp} }  \frac{\partial^2 \phi_E }{\partial t^2} 
\label{SecondOrderCon}
\end{equation}
therefore have to be satisfied. In general, one can expect that only a subset
of all possible boundary values $ \phi_E|_{\partial \Omega_E}$ obey this
condition.  Such a possibility might be attractive because it may put
restrictions on the form of the allowed boundary conditions, based on the
requirement of mathematical consistency.  For the typical equation of motion
under consideration, Eq.~(\ref{Wave}), the requirement (\ref{SecondOrderCon})
leads to the following condition of continuity of the source term
\begin{equation}
\lim_{t\rightarrow \pm {t_{\rm max}}^{\pm}}S[\phi_{L\pm}]= 
\lim_{t\rightarrow \pm {t_{\rm max}}^{\mp} }S[\phi_E]\, . 
\end{equation}  

In order to illustrate some features of the method discussed above, let us
study the equation of motion
\begin{equation}
\frac{\partial^2 \phi}{\partial t^2}-\beta(t) \frac{\partial^2\phi}{\partial x^2}=0\, ,  
\label{StepFunctionEq}
\end{equation}
with the step-like $\beta$-function
\begin{equation}
\beta(t) = \left\{   \begin{array}{ccc}   
1  & {\rm for} & t >  t_{\rm max}\, ,   \\  
-1 & {\rm for} &  t_{\rm max} \geq  t \geq   -t_{\rm max}\, ,  \\ 
1 & {\rm for} &  t < -t_{\rm max}\, .
\end{array} \right.
\end{equation}
In this (1+1)-dimensional example the variation of $\beta$ is not continuous
across the signature change, therefore condition (\ref{SecondOrderCon}) will
have no chance of being fulfilled. Nevertheless, let us choose the following
boundary conditions:
\begin{equation}
\phi_{\pm} = c_{\pm} \cos(x)\, , 
\end{equation} 
where $c_{\pm} \in \mathbb{R}$ are constants. (The boundary condition at
$\partial \Omega_{\infty}$ will not play any role.)  With boundary values
specified, a solution to Eq.~(\ref{StepFunctionEq}) in the Euclidean domain
$\Omega_E= \left\{(t,x)\in \mathbb{R}^2, t_{\rm max} \geq t \geq -t_{\rm
    max}\right\}$ can be found:
\begin{equation}
\phi_E = \cos(x) \frac{c_{+}{\rm sh}(t+t_{\rm max})-c_{-}{\rm sh}(t-t_{\rm
    max})}{{\rm sh}(2t_{\rm max})}\, .   
\end{equation}
Using the solution we can find that 
\begin{equation}
\lim_{t\rightarrow \pm {t_{\rm max}}^{\mp} }  \frac{\partial \phi_E }{\partial t}  =
\cos(x) \frac{\mp  c_{\mp}\pm c_{\pm}{\rm ch}(2t_{\rm max})}{\sinh(2t_{\rm max})}\, ,  
\end{equation} 
which will be used to impose Cauchy initial conditions in the Lorentzian
domains.  Taking this into account, solutions to Eq.~(\ref{StepFunctionEq}) in
the domains $\Omega_{\pm}= \left\{(t,x)\in \mathbb{R}^2, \pm t > t_{\rm
    max}\right\}$ can be determined:
\begin{eqnarray}
\phi_{L+} &=& \cos(x) \frac{  c_{+}\left({\rm sh}(2t_{\rm max})\cos(t-t_{\rm max}) 
+{\rm ch}(2t_{\rm max})\sin(t-t_{\rm max})\right)-c_{-}\sin(t-t_{\rm max}) }{{\rm sh}(2t_{\rm max})}\, ,   \nonumber \\ 
&&\\
\phi_{L-} &=& \cos(x) \frac{  c_{-}\left({\rm sh}(2t_{\rm max})\cos(t+t_{\rm max}) 
-{\rm ch}(2t_{\rm max})\sin(t+t_{\rm max})\right)+c_{+}\sin(t+t_{\rm max}) }{{\rm sh}(2t_{\rm max})}\, .   \nonumber \\ 
&&
\end{eqnarray}

\begin{figure}
\begin{center}
\includegraphics[width=12cm]{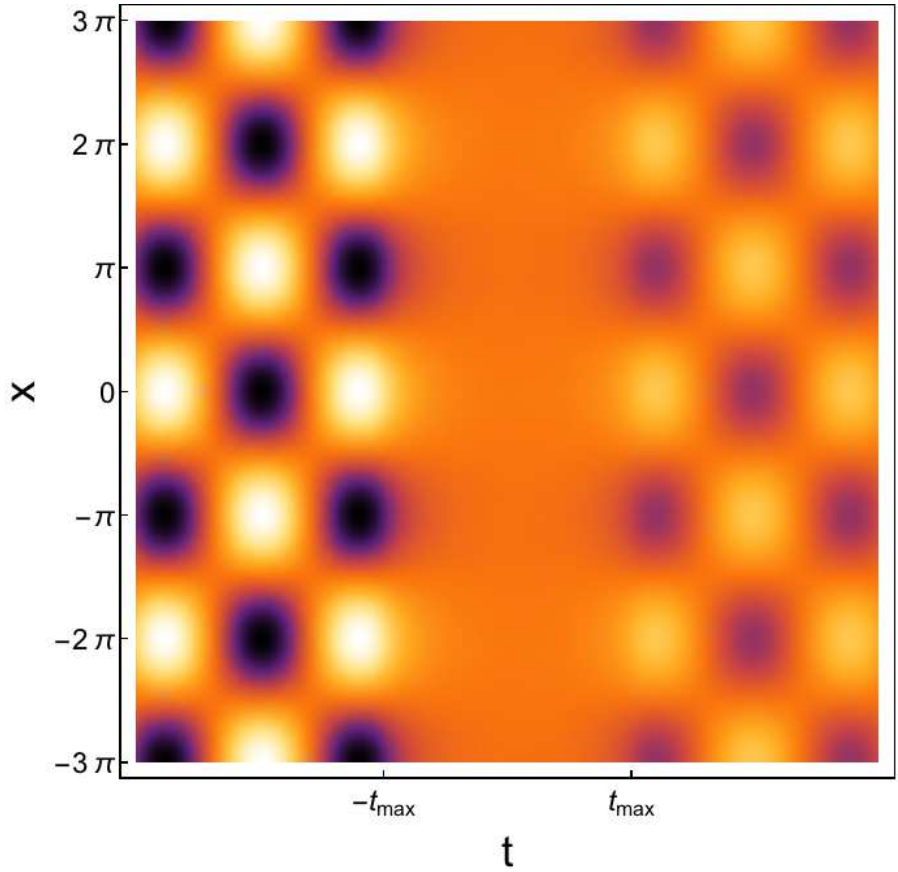}
\caption{ Density plot of the solutions  $\phi_E$ and $\phi_{L\pm}$ to the equation 
(\ref{StepFunctionEq}).  The observed asymmetry is due to the different  boundary 
conditions at $t_{\rm max}$ and $-t_{\rm max}$: $\phi_{\pm}=c_{\pm} \cos (x)$. 
The values $c_{+}=1$, $c_{-}=2$ and $t_{\rm max}=4$ have beed used.}
\label{EuclidSol1}
\end{center}
\end{figure}

The solutions $\phi_E$ and $\phi_{L\pm} $ are presented in
Fig.~\ref{EuclidSol1} in the form of the density plot for the sample values of
$c_{\pm}$ and $t_{\rm max}$.

The solutions obtained in this way are regular in the whole domain
$\Omega=\Omega_E\cup\Omega_{+} \cup \Omega_{-}$. However, in agreement with
the earlier expectations, the second derivatives are discontinuous across the
signature change:
\begin{equation}
\lim_{t\rightarrow \pm {t_{\rm max}}^{\mp} }  \frac{\partial^2 \phi_E }{\partial t^2} 
-\lim_{t\rightarrow \pm {t_{\rm max}}^{\pm}}\frac{\partial^2 \phi_{L\pm} }{\partial t^2}  = 2 c_{\pm} \cos(x)\, .  
\end{equation}

\subsubsection{Wave functions}

Mixed-type differential equations for modes are a consequence of properties of
effective equations in cosmological models of loop quantum gravity. These
properties follow from a general analysis of consistent and anomaly-free
realizations of field equations, so that physical predictions made in these
models are independent of coordinate or gauge choices. The same modifications
to space-time structure can be seen from commutator algebras at the quantum
level, but it is then more difficult to derive dynamical effects. It is
nevertheless interesting to discuss how signature change or its accompanying
instabilities could be seen in the evolution of wave functions.

As long as one has unitary evolution, as can easily be guaranteed in
deparameterized models which provide the majority of explicit wave-function
evolutions in quantum cosmology, the wave function cannot be subject to
instabilities. (However, we note that at a fundamental level it may be
difficult to have unitary evolution under all circumstances, owing to the
problem of time.) Its norm is preserved even throughout high
density. Accordingly, the Schr\"odinger or Klein--Gordon type equation which
the wave function is subject to does not change its form to become elliptic;
the relevant coefficients in the wave equation do not change sign. After all,
what changes is the signature of space(-time), not the signature of
(mini)superspace on which the wave function is defined. (See also
\cite{SigChangeMini}. In fact, depending on which representation one chooses
for the wave function, the equation it is subject to may be elliptic or
hyperbolic even for standard space-time.)

Signature change implies instabilities in an initial-value problem for the
modes, solving effective equations. The fields in effective equations are
expectation values of mode operators in a state obeying the equation for the
wave function. At the level of wave-function evolution, signature change would
therefore be recognized in an exponential change of expectation values of mode
operators. (If wave packets are used, their peak position would change
exponentially.) For these expectation values (or peak positions), the same
sensitivity to initial values as observed in solutions of effective equations
would occur, even if they are computed from an evolved wave function. These
observable quantities therefore evolve in an unstable way, causing the
same problem as seen in an initial-value formulation for effective
equations. Instability problems are just more hidden in schemes that first
evolve wave functions and then compute expectation values, but they are just
as pressing as in effective equations. 

In this respect, the situation is reminiscent of quantum chaos, which,
compared with the classical phenomenon, is more difficult to define and
discuss, but not absent, for wave functions subject to linear differential
equations. In both cases, the sensitivity of evolved expectation values to
initial choices is relevant. It has been discussed in detail in the context of
quantum chaos. Following \cite{StabilityQuantumChaos,EnvQuantumChaos}, we can
use the same arguments for states in the presence of signature change: Unitary
quantum evolution ensures that the overlap
$|\langle\psi_1(t),\psi_2(t)\rangle|^2$ of two different initial states
$\psi_1(t_0)$ and $\psi_2(t_0)$ is conserved in time, and therefore does not
indicate any sensitivity to choices of initial values. However, quantum
evolution of classically chaotic systems is very sensitive to perturbations of
the Hamiltonian operator, in the sense that the overlap
$|\langle\psi_1(t),\psi_2(t)\rangle|^2$ changes rapidly if $\psi_1(t)$ and
$\psi_2(t)$ are now defined as states evolved from the same initial wave
function $\psi_1(t_0)=\psi_2(t_0)$ but with different Hamiltonians
$\hat{H}_1=\hat{H}$ and $\hat{H}_2=\hat{H}+\epsilon\hat{V}$ with some small
$\epsilon$ and a perturbation potential $\hat{V}$. Similarly, the same
definition of the overlap in Euclidean signature leads to rapid (exponential)
change because the perturbed evolution by $\hat{H}_2$ contains a factor of
$\exp(\epsilon\hbar^{-1} \hat{V} t)$. In discussions of quantum chaos,
$\epsilon \hat{V}$ is usually thought of as coming from unknown interactions
with an environment hard to control. The same source of $\epsilon \hat{V}$
exists in quantum cosmology because the precise degrees of freedom and
interactions at high density are not well known. In addition to this, loop
quantum gravity is subject to a large set of quantization ambiguities, so that
the dynamics is not precisely determined and a second source of
$\epsilon\hat{V}$ results.

Signature change in effective equations, in all existing models, is implied by
the requirement of anomaly freedom, so that predictions are guaranteed to be
independent of gauge choices. Most models in which one evolves wave functions
are based on deparameterization, in which one formulates ``evolution'' of a
wave packet in terms of a so-called internal time, which is not a coordinate
but one of the dynamical fields.  The conditions on well-defined evolution
after deparameterization are less severe than the general covariance
conditions when one does not select a specific time, be it a coordinate or
internal. If one were to fix the gauge and work in one specific set of
coordinates, one could avoid having to use effective equations subject to
signature change. Similarly, the possibility of formulating stable
wave-function dynamics in one chosen internal time does not mean that wave
functions in general evolve in a stable manner. One would have to show, first,
that predictions of one's model do not depend on which field is chosen as
internal time (a task which has not been performed in deparameterized models
of quantum cosmology; see \cite{ReducedKasner,MultChoice}). After this, one
could reliably analyze evolution and test whether it is always meaningful or
has to be stopped when a Euclidean regime is reached. Performing this task
turns out to be much more complicated than analyzing anomaly freedom of
effective equations. But interestingly, even in the absence of such an
analysis there are hints that wave-function evolution becomes unstable at high
density, in the sense that expectation values of modes change exponentially
\cite{InhomThroughBounce}.

\subsubsection{Cosmological implications}

A well-posed treatment of initial and boundary values in loop quantum
cosmology implies significant departures from the scenarios commonly made in
this setting. For instance, there has been some interest in a
``super-inflationary'' phase at high density, around a bounce, during which
the Hubble parameter grows quickly \cite{SuperInflLQC}. Although the rapid
change happens for background variables and is therefore not the same as the
instabilities of mode equations such as (\ref{Wave}), one can easily check in
explicit models that the super-inflationary phase falls within the elliptic
range of the partial differential equations for inhomogeneity
\cite{Action}. The rapid growth of background variables cannot be used for
cosmological effects because their values at the border of the Euclidean phase
must be prescribed as part of the boundary-value problem. Obtaining a large
parameter with an ill-posed initial-value problem does not have physical
significance.

A related effect has been seen in a combination of loop-modified background
equations with inflationary models: At high density, the modified equation for
the inflaton has an anti-friction term which can easily push the inflaton up
to a high value in its potential \cite{Inflation,BounceCMB}. Given the right
potential for suitable inflation, the correct initial conditions can therefore
be provided. Also this effect is subject to the verdict of being based on
ill-posed data. An alteration may, however, still be realized if one can show
that for a given background value $\bar{\phi}$ at the beginning of the
expansion phase there must generically be a large $\bar{\phi}'$ according to
the well-posed characteristic formulation. Values of $\bar{\phi}$ far from the
minimum of a potential could then be achieved afterwards, using well-posed
evolution with the predicted initial values in the expansion phase.

With the discussion of initial and boundary values in Sec.~\ref{s:Char}, we
can make our expectations for cosmological scenarios more precise. Some data
must be chosen at the beginning of the expansion phase, even if the energy
density or other coefficients of the differential equation never become
infinite. This feature is shared with the singular big-bang model, in which
the main conceptual problem is caused not so much by divergences but rather by
the presence of unrestricted initial data. (In general relativity, it is
possible to extend solutions across singularities in a distributional sense,
but the extension is not unique \cite{HawkingEllis}.) If there would be a
unique way of deriving initial values at the beginning of the expansion phase,
the singularity of the traditional big-bang model would not be so much of a
problem. One could make clear predictions about the initial state and further
evolution of the universe, as attempted in inflationary scenarios with their
(somewhat controversial) initial conditions for the inflaton. In practical
terms, the singularity problem is therefore one of indeterminedness, which is
implied but not necessarily equivalent to the occurrence of divergences.

An example for the latter part of the preceding statement is given by
signature-change cosmology. There are no divergences in these models, and yet
an important part of initial values for the expansion phase remains free. The
resulting scenario, which we call a finite beginning to emphasize the absence
of divergences together with the requirement of choosing new initial values,
is rather close to the standard big-bang model. In particular, for practical
purposes the need for new initial values once the density is sufficiently low
to trigger signature change back to a Lorentzian structure makes the Euclidean
region appear as a singularity. We therefore speak of a finite beginning
instead of a non-singular one, keeping in mind that the main problem of a
singularity is the indeterminedness of initial data.

There may yet be a crucial difference with the standard big-bang model: What
we need to prescribe at the finite beginning, according to Fig.~\ref{Fig:IB},
is only the field $\phi$, not its normal derivative $\phi'$ if we start
evolution in the collapse phase. If one has some means to know $\phi$ in a
neighborhood of the border between the Euclidean region and the expansion
phase, for instance from some hypothetical observations, one can derive
$\phi'$ and draw conclusions about the field in the collapse phase. There is
therefore some connection between collapse and expansion, although it is not
causal and not fully deterministic. A thorough cosmological analysis of scalar
and tensor power spectra in signature-change models is required before one can
tell how much about the collapse phase could be deciphered in this indirect
way.

\subsubsection{Global issues}
\label{s:Global}

While effective methods for constrained systems provide reliable local
equations, as confirmed by the results of this paper, there may seem to be
several global problems related to the new type of partial differential
equations which imply the disappearance of time or causality at high
density.\footnote{``And well I know it is not right// to seek and stay Time in
  his flight.'' \cite{Don}} One example has been discussed in \cite{Loss} in
the context of black holes, where the non-singular beginning in cosmology as
detailed here is replaced by a naked singularity (again in the sense of
indeterminism) with a Cauchy horizon. Even if local field equations are
regular, the lack of a causal structure in some regions may lead to
unacceptable indeterministic behavior at a global level.

A different kind of global problem is indicated by one feature of solutions to
Tricomi's problem which we have not mentioned so far. Again, locally the
equation and its solutions are well-defined. But generic solutions turn out to
have a pole at one point $A\cap B$ in Fig.~\ref{Fig:Tricomi}, where the
Euclidean arc ends \cite{Tricomi}: Although solutions are smooth in the
interior of the characteristic region (for smooth boundary data) most of them
have a pole at one endpoint of the boundary. (This result explains the
ubiquity of sonic booms in analogous acoustic models.) The solution may remain
finite, but not all of its derivatives do. Derived quantities such as the
energy density in a matter field could therefore diverge. If this happens, it
is not likely that cosmological perturbation theory gives reliable results,
even though the local perturbative equations are regular.

The acoustic analog illustrates a further point: One could think that
signature change in cosmology (or black-hole physics) is harmless because
analogous effects can appear in well-known systems such as transonic motion
(or the model discussed in Sec.~\ref{s:Analog}). The difference is that
transonic motion leads to signature change only for excitations in the fluid,
while all other propagating degrees of freedom including the bulk fluid motion
and space-time are still governed by deterministic equations. (The speed of
light is an absolute limit for any causal motion, very much unlike the speed
of sound in a fluid. If $\beta<0$ in (\ref{Wave}), all motion is eliminated.)
If the bulk of the fluid is forced to move faster than its own speed of sound,
it overtakes any sound wave generated in it, so that its density profile
provides future conditions for the wave. Mathematically, this is represented
by Tricomi's future data. However, the bulk fluid itself moves in a
deterministic (but not wave-like) way even if it is faster than its own speed
of sound, and no causality issues appear. The situation is very different when
signature change happens for space-time physics, in which case no reference
time remains to define a causal structure and no mode can evolve
deterministically. For this reason, signature change in effective space-time
models of loop quantum gravity has much more radical consequences than
analogous effects in matter systems, as discussed in the present paper as well
as \cite{Loss}.

\section{Conclusions}

We have described several fundamental properties of a new scenario, based on
signature change, that has emerged from spherically symmetric and cosmological
models of loop quantum gravity in recent years. We emphasize that this
scenario cannot be seen in the minisuperspace models traditionally studied in
loop quantum cosmology \cite{LivRev}. In fact, the possibility of signature
change casts significant doubt on the viability of minisuperspace models of
loop quantum cosmlogy because in such models one would not see any sign of a
disappearing causal structure at high density. Minisuperspace models of loop
quantum cosmology may be used for some estimates of background properties, but
they can no longer be considered as reliable sources of stand-alone
cosmological scenarios. One always has to go beyond homogeneity to make sure
that there is a well-defined space-time structure and to check whether mode
equations remain hyperbolic, or to exclude other exotic effects.

Signature-change cosmology is therefore a new scenario which crucially relies
on inhomogeneous features of loop quantum gravity (even though, needless to
say, it has not yet been derived from a full quantization of gravity). Its
details, regarding for instance power spectra, still have to be worked out,
but we believe that the more mathematical and conceptual properties discussed
in this article already show a large number of interesting features. The main
consequence in practical terms is the occurrence of instabilities, related to
the sensitive dependence of solutions on initial data. In some examples,
instabilities may be physical effects which imply rapid change but no
inconsistencies. In quantum gravity, however, instabilities of the kind
encountered in the presence of signature change are fatal: The theory remains
subject to a large number of quantization ambiguities in its equations, and
not much is known about suitable initial states for quantum space (and not
just quantum matter). With this inherent vagueness, one cannot afford any
instabilities that would magnify theoretical uncertainties in a short amount
of time, even if this time may be as small as the Planck time. No predictions
would be possible. We are therefore sympathetic with the verdict ``In light of
the fact that even this 'well-behaved' signature change system predicts its
own downfall, it may be prudent to reassess the inclusion of signature
changing metrics in quantum gravity theories.''  \cite{KleinSigChange}
obtained after a detailed analysis of initial-value problems in a model of
classical signature change. Our model is provided by effective equations of
loop quantum gravity in which signature change appears to be generic (and is
not included by choice). The verdict of \cite{KleinSigChange} can therefore be
applied to loop quantum gravity at least to the extent that extreme caution is
called for when one considers evolution through high density.

In the presence of signature change, instabilities can be avoided only if one
switches to a 4-dimensional boundary-value problem at high density, giving up
causality. Although ambiguities remain in the theory, there are several
qualitative effects with interesting implications. As one of the main
observations in this article, the mixed-type partial differential equations
for modes in this context strike a nice balance between deterministic cyclic
models and singular big-bang models. There are no divergences, and yet initial
data in the infinite past do not uniquely determine all of space(-time). For
every mode, one must specify one function at the beginning of the expansion
phase even if one has already chosen initial values for the contraction
phase. Still, the normal derivative of the field is not free and may carry
subtle but interesting information about the pre-big bang.

\section*{Acknowledgements}

This work was supported in part by NSF grant PHY-1307408 to MB. JM is
supported by the Grant DEC-2014/13/D/ST2/01895 of the National Centre of
Science.

\begin{appendix}

\section{Space-time}
\label{s:SpaceTime}

We present here a somewhat technical discussion of fundamental properties of
space-times underlying (\ref{Wave}). Symmetries and gauge transformations are
especially important in this context, as well as properties of the canonical
formulation of gravity. In our following exposition, we also provide more
details on the derivation and reliability of equations such as (\ref{Wave}) in
effective models of loop quantum gravity.

\subsection{Gauge transformations}

In spite of its appearance, equation (\ref{Wave}) is covariant in a
generalized sense, according to an effective (and canonically defined)
non-Riemannian structure of quantum space-time. The usual Lorentz and
Poincar\'e symmetries, under which the classical version of (\ref{Wave}) with
$\beta=1$ (and constant $a$) would be invariant, are realized in canonical
gravity as a subalgebra of the infinite-dimensional algebra (or rather,
algebroid \cite{ConsAlgebroid}) of deformations of 3-dimensional spacelike
hypersurfaces in space-time. (See for instance \cite{DeformedRel,CUP}.)  These
hypersurface deformations are gauge transformations of any generally covariant
theory, including gravity.

Hypersurface deformations \cite{Regained} are more suitable than Poincar\'e
transformations in situations in which no background space-time metric is
assumed. They provide the proper framework for a discussion of generalized
space-time structures as the may be implied by canonical quantum
gravity. Hypersurface deformations have an infinite set of generators, spatial
ones given by $D[N^a]$ with spatial vector fields $N^a$ tangential to the
hypersurface, and normal ones $H[N]$ with functions $N$ on the hypersurface so
that the deformation is by an amount $Nn^a$ along the unit normal $n^a$. The
classical space-time geometry \cite{Regained} (as well as the canonical form
of general relativity \cite{DiracHamGR}) imply that these generators have
commutators (or classical Poisson brackets)
\begin{eqnarray}
 \{D[N_1^a],D[N_2^b]\} &=& D[N_1^a{\rm D}_bN_2^b-N_2^a{\rm D}_aN_1^b]\\
 \{H[N],D[N^a]\} &=& H[N^a{\rm D}_aN]\\
 \{H[N_1],H[N_2]\} &=& -D[h^{ab}(N_1{\rm D}_bN_2-N_2{\rm D}_bN_1)] \label{HH}
\end{eqnarray}
with the induced metric $h_{ab}$ and covariant derivative ${\rm D}_a$ on a
hypersurface.  

When one tries to quantize the theory canonically, one should turn the gauge
generators into operators, so that they still have closed brackets given by
commutators. Otherwise, the classical gauge transformations would be broken
and the quantum theory would not be consistent; it would have gauge anomalies.
The anomaly problem of canonical quantum gravity is important, but also very
difficult and unresolved so far. (One reason is the presence of the metric in
(\ref{HH}), which at the quantum level would be an operator and give rise to
complicated ordering questions with a quantized $D[N^a]$.) Nevertheless, there
have been several independent indications in recent years which give some hope
that the problem can be solved. At the operator level, especially
$2+1$-dimensional models have been analyzed in quite some detail, paying
attention to the full algebra of gauge generators. Consistent versions have
been found in different ways
\cite{ThreeDeform,TwoPlusOneDef,TwoPlusOneDef2,AnoFreeWeak}, including also
spherically symmetric models \cite{SphSymmOp}.

Independently, effective calculations start from the observation that the
quantum operation of commutators together with a closed algebra of operators
$\hat{C}_I$, $[\hat{C}_I,\hat{C}_J]=\hat{f}^K_{IJ}\hat{C}_K$ with structure
constants (or functions/operators) $\hat{f}^K_{IJ}$, implies a closed algebra
under Poisson brackets of effective constraints
$\langle\hat{C}_I\rangle$. These effective constraints are defined as
expectation values $\langle\hat{C}_I\rangle$ of the constraint operators in an
arbitrary state \cite{EffCons,EffConsRel}. Effective constraints are therefore
functions on the quantum state space, which can conveniently be parameterized
by expectation values and moments with respect to a set of basic operators. In
addition to these effective constraints, obtained as direct expectation values
of constraint operators, there is an infinite set of independent ones, derived
from the same constraint operators as $\langle\widehat{\rm
  pol}\hat{C}_I\rangle$ for all polynomials in basic operators. All these
functions on the space of states would be zero on the subspace annihilated by
the constraints $\hat{C}_I$, imposed following Dirac's prescription.  The need
for an infinite set of effective constraints for every single constraint
operator follows from the requirement that a whole tower of infinitely many
moments must be constrained together with every constrained expectation value.

A Poisson bracket on the quantum state space of expectation values and moments
can be defined by $\{\langle\hat{A}\rangle,\langle\hat{B}\rangle\}=
\langle[\hat{A},\hat{B}]\rangle/i\hbar$, extended by the Leibniz rule to
products of expectation values (as they appear in quantum fluctuations and
higher moments). With these definitions, it follows that effective constraints
form a closed algebra under Poisson brackets if the constraint operators form
a closed algebra under commutators. Practically, it is easier to evaluate
Poisson brackets than commutators, an observation on which the idea of
canonical effective equations \cite{EffAc,Karpacz} and constraints
\cite{EffCons,EffConsRel} is based. Also, there are useful approximation
schemes, such as a semiclassical one in which one would do calculations order
by order in the moments, which are easier to implement for effective
constraints than for constraint operators and allow one to handle the unwieldy
space of all states more efficiently. For finite orders in the moments, there
is a finite number of independent effective constraints, and the task of
computing their algebra becomes feasible.

If we have a closed algebra $[\hat{C}_I,\hat{C}_J]$ of operators, the
corresponding effective constraints truncated at some moment order form a
closed Poisson-bracket algebra up to this order, irrespective of the states
used. This formulation of effective theories is therefore much more general
than the usual idea of effective actions \cite{EffConsQBR}. (The latter are
often combined with further approximations such as derivative expansions, or
with restrictions of states such as near-vacuum states for the low-energy
effective action.) In our case, we need not make any assumption on the class
of states in order to obtain a closed algebra of effective constraints. One
may just have to include higher orders in the moments for a better
approximation to solutions corresponding to strong quantum states; but even at
lower orders, the algebra must be closed (to within the same order). Effective
constraints therefore provide a good test of possible anomaly-free quantum
constraint algebras. In particular, if one can show that no closed effective
constraint algebra exists to within some order in moments, using a
large-enough parameterization of quantum corrections, there cannot be a closed
algebra of constraint operators. And if closure of effective constraints can
be achieved only if the classical constraint algebra is modified, the full
quantum constraint algebra must be subject to quantum corrections which change
the form of gauge transformations. Signature change is just one of the
consequences found in this way: Instead of (\ref{HH}), we then have
\begin{equation} \label{HHbeta}
\{H[N_1],H[N_2]\} = -D[\beta h^{ab}(N_1{\rm D}_bN_2-N_2{\rm D}_bN_1)]
\end{equation}
with a phase-space function $\beta$ which may turn negative (while the other
brackets involving $D[N^a]$ remain unchanged). Canonical field equations
consistent with this modified bracket are of the form (\ref{Wave}), as shown
by \cite{Action} using the methods of \cite{Regained,LagrangianRegained}. (The
same modification of derivative terms is realized to higher orders in
derivatives \cite{ActionHigher}.)

So far, calculations in models of loop quantum gravity have been performed
only to zeroth order in the moments. But they still show interesting effects
because, in addition to moment terms, the theory implies further quantum
corrections. At high density, holonomy modifications are relevant, which are
implied by the basic assumption of loop quantum gravity
\cite{ThomasRev,Rov,ALRev} that the gravitational connection can be
represented as an operator only when it is integrated and exponentiated to a
holonomy. A second effect, inverse-triad corrections
\cite{InvScale,QuantCorrPert,LoopMuk}, is more indirect but also related to
the basic assumption. It implies corrections relevant at lower curvature
\cite{LoopMuk,InflTest,InflConsist}. Effective constraint algebras have been
computed in both cases and found to allow for consistent versions:
\cite{ConstraintAlgebra,LTBII,ModCollapse} as examples for inverse-triad
corrections, \cite{ScalarHol} for holonomy corrections, and
\cite{JR,ScalarHolInv,HigherSpatial} for combinations of both. Whenever
holonomy modifications are present, all generic consistent effective
constraints found so far imply mode equations of the form (\ref{Wave}).

The derivation shows that covariance is not broken because all classical gauge
generators have a valid analog as a constraint operator $\hat{C}_I$ or as an
effective constraint. For every classical gauge transformation, there is a
corresponding quantum or effective gauge transformation. No transformations
are violated, including the Lorentz and Poincar\'e ones that one obtains as a
special case of hypersurface deformations \cite{DeformedRel}. Nevertheless, it
is possible for (\ref{Wave}) to differ from standard covariant wave equations
because quantum gauge generators, while they must not be broken for an
anomaly-free theory, may be subject to quantum corrections. These corrections
can be seen in the structure operators or functions $\hat{f}^K_{IJ}$ of the
quantum or effective constraint algebras. For holonomy and inverse-triad
corrections, the classical structure functions are, as in (\ref{HHbeta}),
multiplied with a phase-space function $\beta$, which determines a new
consistent form of quantum space-time covariance. The same function appears in
mode equations (\ref{Wave}) derived from the effective constraints.

\subsection{Slicing independence}

Even though the speed in (\ref{Wave}) provided by quantum geometry depends on
the spatial metric and extrinsic curvature --- quantities that are not
covariant in classical space-time --- the effect is frame independent in a
subtle way. The models in which such modified speeds have been derived are
anomaly free, so that the classical set of gauge transformations (given by
coordinate changes) is not violated. However, not just the dynamics but also
the structure of space-time receives quantum corrections. There is a new set
of gauge transformations under which the quantum-corrected field equations are
invariant, and which has the full set of standard coordinate changes as its
classical limit. The effective theories, including modified speeds they
predict, are covariant under these deformed gauge
transformations. Accordingly, the space-time structure is no longer classical
or Riemannian, but it remains well-defined in canonical terms.

Put differently, one could worry that models in which waves propagate with
speeds depending on the spatial metric and extrinsic curvature of
constant-time hypersurfaces violate the slicing independence of the classical
theory. Classically, for any given space-time, one can, depending on one's set
of coordinates, choose equal-time slices with large or small extrinsic
curvature, even if the covariant curvature of space-time vanishes. A speed
depending on extrinsic curvature would then suggest that predictions depend on
the slicing or the choice of an initial-value surface, which would be
unacceptable. This worry is unjustified because the two initial-value
surfaces, one with small and one with large extrinsic curvature, would give
rise to different physical solutions of the effective theory even though they
would present the same classical solution. Predictions, for instance of the
speed, then differ simply because one would consider physically different
solutions. Transformations between different slicings are, just like
coordinate changes, gauge transformations which are modified in the effective
theory. Slicings of one and the same classical space-time, which are related
by classical gauge transformations, are not gauge related in the
quantum-corrected setting.

\end{appendix}

%\bibliographystyle{../preprint}
%\bibliography{../Bib/QuantGra.bib}

\end{document}